\documentclass{article}

\usepackage{arxiv}

\usepackage[utf8]{inputenc} 
\usepackage[T1]{fontenc}    
\usepackage{hyperref}       
\usepackage{xurl}            
\usepackage{booktabs}       
\usepackage{amsfonts}       
\usepackage{nicefrac}       
\usepackage{microtype}      
\usepackage{lipsum}		
\usepackage{graphicx}
\usepackage{natbib}
\usepackage{doi}

\usepackage{amsmath,amssymb,amsfonts}
\usepackage{algorithmic}
\usepackage{graphicx}
\usepackage{textcomp}
\usepackage{multirow}
\usepackage{multicol}
\usepackage{xcolor}
\usepackage{caption}

\def\BibTeX{{\rm B\kern-.05em{\sc i\kern-.025em b}\kern-.08em
    T\kern-.1667em\lower.7ex\hbox{E}\kern-.125emX}}
    
\title{Accessing Convective Hazards Frequency Shift with Climate Change using Physics~-~Informed~Machine~Learning}

\author{
Mikhail Mozikov, Ilya Makarov, Alexandr Bulkin, Daria Taniushkina, Roland Grinis, and Yury Maximov \\[1ex]
	\texttt{mb.mozikov@gmail.com} 
}

\renewcommand{\shorttitle}{Accessing Convective Hazards Frequency Shift}

\hypersetup{}

\begin{document}
\maketitle

\begin{abstract}
	In this paper we discuss and address the challenges of predicting extreme atmospheric events like intense rainfall, hail, and strong winds. These events can cause significant damage and have become more frequent due to climate change. Integrating climate projections with machine learning techniques helps improve forecasting accuracy and identify regions where these events become most frequent and dangerous. To achieve reliable and accurate prediction, we propose a robust neural network architecture that outperforms multiple baselines in accuracy and reliability. Our physics-informed algorithm heavily utilizes the whole range of problem-specific physics, including a specific set of features and climate projections data.  The analysis also emphasizes the landscape impact on the frequency distribution of these events, providing valuable insights for effective adaptation strategies in response to climate change.
\end{abstract}

\keywords{Atmospheric Event \and Climate Change \and Machine Learning}

\section{Introduction}
\label{sec:introduction}
Predicting the frequency of severe convective storms (SCS) and associated hazards are essential to mitigate their impact, protect people and infrastructure. These storms, often characterized by intense thunderstorms with strong updrafts and downdrafts, can produce tornadoes, damaging winds, hail, and heavy rainfall. 

Accurate convective hazards frequency forecasting of SCS provides valuable insights for infrastructure development, urban planning, and disaster response readiness. Policymakers and city planners can incorporate measures like storm-resistant buildings, efficient drainage systems, and emergency response protocols.

Moreover, forecasting SCS frequency  contributes to understanding climate change impacts on severe weather patterns. Rising global temperatures may alter atmospheric dynamics, potentially affecting the frequency and intensity of convective storms. Studying these changes enhances our comprehension of climate variability, improves climate models, and strengthens our ability to adapt to future challenges.

Finally, accurate forecasting of convective hazardous phenomena associated with severe convective storms is crucial. It facilitates effective preparedness and response, minimizing loss of life and property, and informs long-term planning strategies. Given the ever-changing climate and its potential impacts on severe weather events, continuous research and development in this field are vital for building resilient communities and safeguarding the well-being of individuals and societies.

\noindent{\bf State-of-the-Art.}
Weather and climate forecasts play a crucial role in various industries, including agriculture \cite{mjelde1998review}, transportation \cite{love2010climate}, energy \cite{voisin2006role}, and disaster management \cite{tall2012using}. Different approaches exist for meteorological and climate forecasts, including numerical weather prediction (NWP) \cite{bauer2015quiet}, statistical modeling \cite{gneiting2014probabilistic}, and machine learning (ML) techniques \cite{bochenek2022machine}. NWP, which relies on mathematical models to simulate atmospheric conditions and provide predictions based on global and regional models, is the most commonly used method. Statistical modeling utilizes historical data being extremely useful when data is limited or the atmospheric event is less complex. 

Increasingly popular machine learning (ML) often allows for improving the results against classical statistical approaches.
Ensemble forecasting~\cite{vincendon2011perturbation} involves running multiple NWP models with different conditions to create a range of possible outcomes, while analog forecasting~\cite{chattopadhyay2020analog} looks for historical events with similar weather patterns to make predictions. The integration of ML has led to significant advancements in meteorology, with applications in agro-meteorology~\cite{kelley2020using} and weather phenomena like heavy rains~\cite{choi2018development}. ML techniques aid in estimating crop water demand using on-farm sensors and enhancing climate models. Notable studies involve merging satellite and climate data for wheat yield predictions using ML~\cite{burke2020calibration}.

In~\cite{watson2021machine}, the connection and disparity between ML applications in weather and climate forecasting are explored. Meteorology focuses on short-term predictions, while climate forecasting deals with long-term trends. Due to the distinct timescales, the methods and models used should be chosen carefully and adapted for accurate results. Major ML approaches suitable for both weather and climate forecasting include the use of Visual Transformer-based models~\cite{nguyen2023climax}, Convolutional Neural Networks (CNNs)~\cite{abdalla2021deep}, Recurrent Neural Networks (RNNs)~\cite{zaytar2016sequence}, Generative Adversarial Networks (GANs) \cite{chen2019generative}, Autoencoders, and classic ML approaches~\cite{tibau2021spatio},~\cite{czibula2021autonowp}.

ML approaches have great potential to provide accurate forecasts, but success heavily relies on the data quality and the training features. Although data quality has improved, obtaining datasets with long time spans for specific atmospheric phenomena remains challenging. ML approaches can be computationally expensive and require high-performance computing resources, but this disadvantage is not critical when compared to the computational demands of NWP~\cite{7328658}.

ML algorithms have already helped with many weather and climate related problems, but long-term SCS frequency estimation forecasting still needs more attention. The latter is likely due to the limited availability of high-accuracy datasets and low spatial and time resolutions of reanalysis and climate projection data. An example of ML application is given in \cite{zhou2019forecasting}, where the authors present a deep learning forecasting solution for severe convective weather, including short-duration heavy rain, hail, convective gusts, and thunderstorms based on numerical weather prediction data with a 12-hour lead time. The scope of this research is focused on China. Research ~\cite{czernecki2019application} combines real-time radar reflectivity, lightning frequency, and reanalysis data sets to predict significant hail events. 

Whereas previous papers have offered relevant insights into various weather and climate forecasting aspects, our study seeks to contribute to understanding long-term SCS frequency and forecasting. By leveraging ML techniques and the unique capabilities of CMIP climate projections, we aim to provide valuable insights into this underexplored area of research. 


\section{Problem Setup and Approach}
We aim to design a machine learning model for predicting the changes in SCS activity within a specific region. Our approach incorporates problem-specific physics with deep neural network approach techniques. The latter allows for substantial improvement of prediction accuracy over limited data and boosts machine learning approach robustness.

\subsection{Data}
\label{sec:data}

In our study, we primarily focus on the European region of Russia in our research. The data we use for our SCS frequency forecasts consists of geospatial time series. It is important to note that all SCSs are connected with cumulus congestus or cumulonimbus clouds, which have an extended vertical structure reaching heights of over 18 km in some cases \cite{Krauss2007}. To address this, we utilize data from different isobaric surfaces, essentially different altitudes, resulting in a three-dimensional spatial structure of the data.

Our climate data results from the Coupled Model Intercomparison Project Phase 5 (CMIP5) output \cite{taylor2012overview}. Various universities and research entities across the globe develop these models based on different assumptions about climate change scenarios. These scenarios range from adopting environmentally friendly practices, also known as the ``green road'', to the opposite extreme of a continuous increase in atmospheric emissions. The assumptions are categorized into four representative concentration pathways (RCPs):
\begin{itemize}
    \item RCP2.6 corresponds to a peak radiative forcing of approximately 3 W/m$^2$ around the middle of the century, followed by a decline to 2.6 W/m$^2$ by 2100.
    \item RCP4.5 represents a stabilization (without overshoot) of radiative forcing at 4.5 W/m$^2$ after 2100.
    \item RCP6.0 represents a stabilization (without overshoot) of radiative forcing at 6 W/m$^2$ after 2100.
    \item RCP8.5 signifies an increase in radiative forcing up to 8.5 W/m$^2$ by 2100.
\end{itemize}

For our study, we select the Scenario Representative Concentration Pathway 8.5 (RCP 8.5), which is considered a "worst-case" option in CMIP5. Recent events worldwide, such as the EU energy U-turn\footnote{\url{https://www.dw.com/en/germanys-energy-u-turn-coal-instead-of-gas/a-62709160}}\footnote{\url{https://www.lemonde.fr/en/economy/article/2022/09/02/despite-climate-commitments-the-eu-is-going-back-to-coal\_5995594\_19.html}} and increased coal consumption in China\footnote{\url{https://www.iea.org/countries/china}}, support the latter or even worse emissions scenario. 

It is important to note that a significant portion of SCSs are typically associated with local weather patterns. We utilize data with the highest possible spatial resolution to avoid losing valuable information due to low spatial density. For this purpose, we employ the simulation provided by the Meteorological Research Institute (MRI-CGCM3) in CMIP5 \cite{yukimoto2012new}. This data source offers one of the highest spatial resolutions of $1^{\circ}$ x $1^{\circ}$, preserving fine-scale details necessary for SCS-related analysis.

In our analysis, we choose specific humidity, temperature, and the two components of the wind vector on 12 isobaric surfaces as the features for training our models. These parameters are available with a time resolution of 6 hours. 

The data for the target variable are obtained from \cite{meteoru}, which provides the raw data in the form of telegrams using the SYNOP format (precisely the KN-01 format for Russian localization). It is essential to mention that this data is only available for locations where meteorological stations are situated. On the other hand, the output data from the CMIP5 models is presented in a regular grid format. To ensure compatibility between the target data and the model input, we extract only the grid cells that contain meteorological stations from the full simulation output during the preprocessing stage. 

\begin{table}[htbp]
\caption{The description of the datasets we used. MRI-CGCM3 RCP8.5 dataset is used to derive features and Rosgidromet data for the target labels.}
\centering
\begin{tabular}{p{195pt}|p{105pt}|p{75pt}}

Dataset & Variables & {Time span} \\
\hline
{Rosgidromet meteorological observations \cite{meteoru}} & Convective Hazardous Phenomena & 2009-2021\\
\hline
\multirow{4}{*}{MRI-CGCM3 RCP8.5 \cite{yukimoto2012new}} & U-component of wind & \multirow{4}{*}{2009-2100}\\
 & V-component of wind &  \\
 & Specific humidity &  \\
 & Temperature &  \\
\hline
\end{tabular}
\label{tab: data-list}
\end{table}

\begin{figure}
\centering
\includegraphics[width = 7cm]{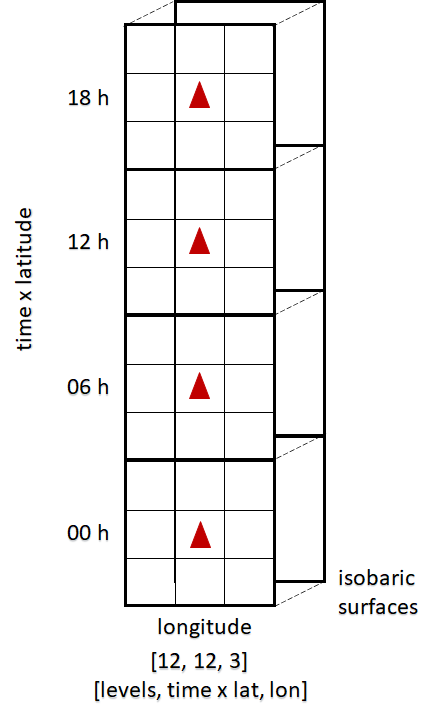}
\caption{Input format. Gridded CMIP5 data are transformed into a form suitable for the model architecture. 3 x 3 cell blocks are cut and stacked along the time axis. Red triangles stand for the location of the same station across different timestamps. Dimension levels represent 12 isobaric surfaces in use.}
\label{fig:input_format}
\end{figure}

For the convenience of data transmission to the convolutional layer, one observation for one feature is given as [levels, time x lat, lon] (Figure \ref{fig:input_format}), where:

\begin{itemize}
    \item levels = 12 - selected isobaric surfaces acting as channels in image processing; 
    \item time x latitude = (4 x 3) - features corresponding to one day are concatenated along the time axis;
    \item lon = 3 - the longitude axis.
\end{itemize}

In the figure , the station (represented by a red triangle) is in the center of the 3 x 3 square for the corresponding observation period, and the surrounding cells are adjacent in the regular grid of dataset points.

For a CatBoost baseline solution, a single observation is presented as a one-dimensional vector of size [layers x latitude x longitude x time].

\subsection{Problem Statement}
We aim to design a machine learning model for predicting the changes in SCS activity within a specific region. 

In our approach, we rely on CMIP5 climate projections and consider them representative of the "real environment" in which we estimate SCS monthly frequency.

We refer to the latter frequency as a ratio of the days in the month when SCS was observed to the total number of days in the given month.
Since our objective is not to forecast SCS events for specific days several years in advance but rather to estimate monthly SCS frequency based on the CMIP5 scenario, we adopt an approach that differs from classic models designed for forecasting. We propose the following algorithm:

\begin{itemize}
    \item Solve a binary classification task for each day to determine whether SCS occurs or not.
    \item Average the obtained results for each month, aggregating the daily forecasts.
\end{itemize}

By employing this algorithm, we can estimate the monthly averaged frequencies of SCS based on the CMIP5 climate projections. This approach allows us to focus on the broader changes in SCS frequency over time rather than attempting to forecast individual SCS events.

In the initial step of our algorithm, we perform a binary classification for each grid point that contains a meteorological station. Class 1 represents the presence of SCS, while Class 0 indicates their absence.

To formalize the optimization problem, we aim to minimize the Binary Cross Entropy (BCE) between the output distributions of our model and the real target values. 

At this step, it is crucial to address the issue of class imbalance, given that Class 1 to Class 0 \ref{f: BCE} ratio is approximately 1:3. To counter this fact, we introduce a constant weight $\lambda$, which we apply to Class 1 before calculating the mean reduction of BCE Loss:
\begin{equation}
    \ell(p, y) = -\frac{1}{N}\sum\limits_{n = 1}^{N}{\left(\lambda y\log(p) + (1 - y)\log(1 - p)\right)},
    \label{f: BCE}
\end{equation}
\noindent where $n \in \mathbb{N}$ is an index of the meteorological station; $y \in \{0; 1\}$ is a true label where class 1 corresponds to SCS; $p$ is a predicted probability of the object being class 1, and $\lambda$ is positive class weight.

By incorporating this weighting scheme, we can appropriately address the class imbalance and ensure our model is not biased toward the majority class. In our experiment, we used the grid search to find the optimal value of $\lambda$, which equals two.

\subsection{Quality Metrics}

We use common classification metrics during training, which include precision, recall, and F1 score. However, after monthly aggregation, it is possible to use regression metrics for vectors of the true and predicted monthly mean values of SCS frequency. At this stage, we estimate our model's performance using standard RMSE and metric, introduced in~\cite{PREIN201810}. 

The referred paper introduces a model designed to hail risk estimation. 
Authors claim they use the specifically normalized version of RMSE to minimize the impact of undersampled SCS observations on the statistic and assess if the algorithm can reproduce the shape of the observed annual cycle, which is less affected by under-observing and artificial trends. 
Since hail is an essential part of SCS hazards and it is also important to understand how well our model gets the annual cycle of SCS, we decided to use the metric in our research. 

Although the exact formula presented in section 4.1 of the referred paper presumably has a typo because it allows for a negative value under the square root. Therefore, the metric used for our model quality estimation is corrected according to the text description presented in the paper. (Formula \ref{f: RMSEac}, original notations preserved).



\begin{equation}
    RMSE_{AC} = \sqrt{\frac{1}{N} \sum\limits_{n = 1}^{N}\|O - P\|^2},
    \label{f: RMSEac}
\end{equation}
where $P$ and $O$ are normalized predicted and observed monthly SCS hazard probabilities over $N$ meteorological stations.
Also, we use a simple Root Mean Square Error measure:
\begin{equation}
    RMSE = \sqrt{\frac{1}{N} \sum\limits_{n = 1}^{N}\|X - X^{pred}\|^2},
    \label{f: RMSE}
\end{equation}
where $X$ and $X^{pred}$ are observed and predicted monthly SCS hazard probabilities, respectively.

\section{Proposed methods}
\subsection{Baseline approach}

At the outset of our experiment, we developed a simple baseline solution to which we could refer. Within this framework, we group all features into a one-day time scale and assume the absence of temporal and spatial structure in it. This assumption is based on the fact that SCS events are typically quite local, and observing a 24-hour period should yield all useful features to detect a single event.

Notably, ignoring spatial structure prevents the model from incorporating information on associated synoptic processes. However, due to the uncertainty of whether, data with a resolution of $1^{\circ} \times 1^{\circ}$ adequately preserves small-scale spatial features corresponding to the SCS formation process, thus the decision to ignore spatial structure was deemed acceptable.

To create the baseline model, we employed the CatBoost classifier. In the subsequent part of this section, we will compare the results of this approach to our neural network solution.

\subsection{Neural network approach}

\begin{figure}[t!]
    \centering
    \includegraphics[width=\linewidth]{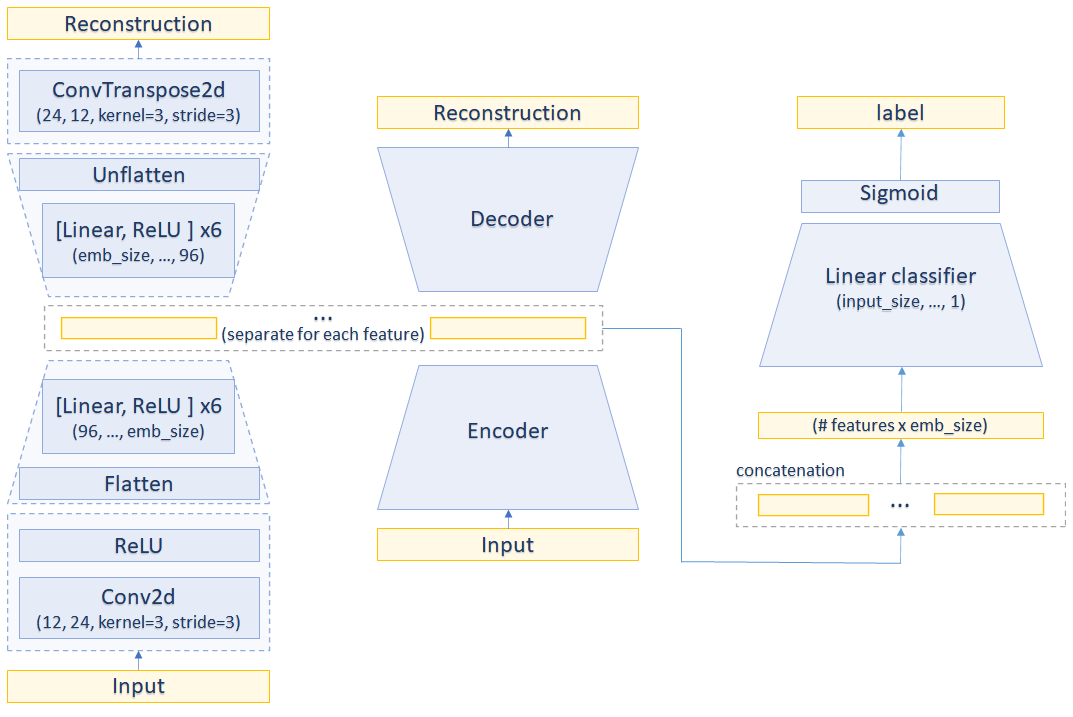}
    \caption{The basic model architecture consists of stacked autoencoders (from left to middle block) pretrained separately for each feature and the classification head (right block) added in the fine-tuning stage.}
    \label{fig: arch}
\end{figure}

The main building blocks of architecture in use are separate pre-trained autoencoders for each of the four features and a linear classifier with resulting embeddings as input. It is possible to use those blocks separately. However, a more convenient approach is to add a classification head and perform autoencoder fine-tuning and linear classifier training processes in parallel. The architecture is given in Figure {\ref{fig: arch}}. This approach already shows significant improvement compared to the CatBoost baseline.  

In this setting, the neural network consists of two main parts: a feature extractor and a classification head. The feature extractor comprises four linear autoencoders with a single convolutional layer at the beginning. This layer helps to extract spatial information, and since we only take a $3~\times~3$ cells block around each station, there is no need for additional convolutional layers. As we mention in Section~\ref{sec:data}, grid parts corresponding to a 24-hour period are stacked along the time axis during preprocessing, so we use a stride equal to 3 for the convolutional layer to prevent data mixing along the time axis. The embeddings are concatenated and forwarded to the classification head at each iteration.

The loss function for this end-to-end training is a sum of the autoencoder reconstruction loss and the classification loss. For the reconstruction, we use Mean Squared Error Loss \ref{f: MSE}:

\begin{equation}
    MSE = \frac{1}{N} \sum\limits_{t = 1 }^{N}\|f(w, X_t) - Y_t\|^2,
    \label{f: MSE}
\end{equation}
where $n \in \mathbb{N}$ is the index of the meteorological station; $X_{t} \in \mathbb{R}^{\text{levels} \times \text{(time x lat)} \times \text{lon}} $ is a tensor of climatic variables at one timestamp;
$Y_{t} \in \{0; 1\}^{\text{time}} $ is a target vector; and $f(w, X)$ is a predicting model.

Stacked hourglass networks inspired the next idea for improvement \cite{newell2016stacked}. Initially used for pose estimation, the hourglass module is shown to capture information at multiple scales. But this feature does not apply to our model, which operates on small spatial dimensions without much scale variability. Nonetheless, we believe this approach can enhance the quality of the model even in different problem statements, as it resembles a well-known skip-connection technique \cite{he2016deep} within our model architecture.

To incorporate this approach, we modify the current model by sequentially stacking three autoencoders for each of the four features. Since our modification differs from the set of hourglass modules, we will refer to it simply as the stacked autoencoders model. 



After forwarding data through autoencoders, we stack all three reconstructions, so we need a way of data aggregation to get the initial dimension and calculate the loss. 
To preserve spatial information in the data, we propose using a 3D convolutional layer to obtain the final reconstruction. 

During pretraining, we observed that the autoencoders for atmospheric temperature and specific air humidity did not show any decline in the reconstruction loss. However, the autoencoders for the wind components demonstrated improvement. 

\section{Experimental Results}
\label{sec: exp_results}

\begin{figure}[htbp]
\centerline{\includegraphics[width=9cm,height=9cm,keepaspectratio]{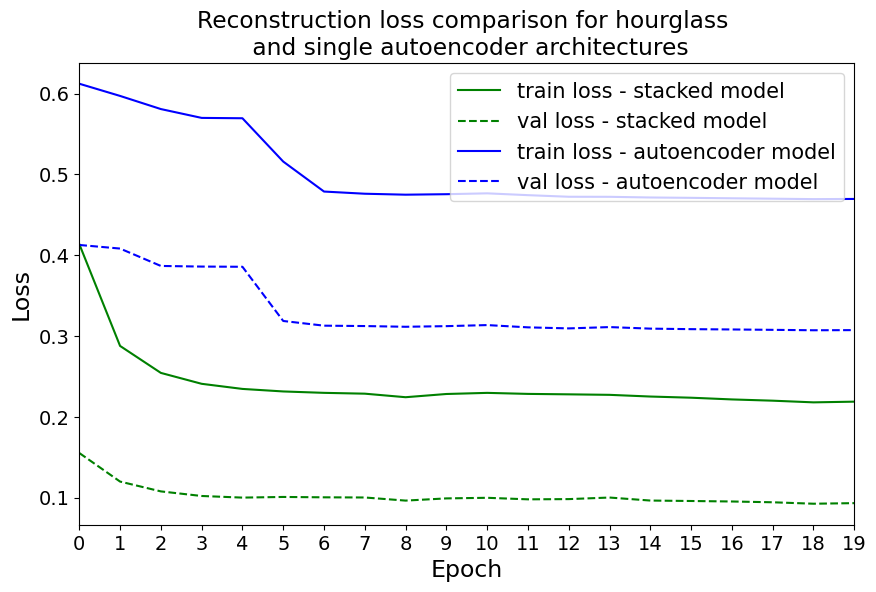}}
\caption{Comparison of the reconstruction loss for different approaches of aggregation. The hourglass architecture (in green) outperforms the single-autoencoder architecture (in blue).}
\label{fig: reconstruction_loss}
\end{figure}

Our experimental analysis reveals that stacking autoencoders produces a notable decrease in the reconstruction loss compared to the straightforward summation technique; see Fig.~\ref{fig: reconstruction_loss} for the details.

Additionally, we have tried three approaches for the embeddings: summation, pixel-wise 1D convolutional, and concatenation followed by a linear layer. 

Table \ref{tab: metrics} summarizes the calculated metrics for all the considered models. Since the original paper \cite{PREIN201810} did not provide open-source code, we took RMSE$_{AC}$ value from the paper.
This table shows that our two-part neural network has already surpassed the CatBoost baseline in terms of performance. As we move towards the first end-to-end model, there is a gradual improvement in metrics. Notably, the Autoencoders \& classifier (end-to-end) model performs best regarding the RMSE$_{AC}$ metric. This indicates that this particular model is the most effective at capturing the normalized annual cycle of SCS activity, while more complex models excel in classification metrics.

In order to minimize potential losses for companies resulting from SCS events, it is advisable to prioritize the maximum recall for Class 1 (SCS) during the classification step while also maintaining high performance in other metrics. Based on our analysis, the Stacked autoencoders with conv1d layer \& classifier model is the most suitable for the problem considered in the paper.

\begin{table*}[htbp]
\renewcommand{\arraystretch}{1.3}
\caption{Models evaluation. Columns 2 to 7 show metrics for the classification stage. High metrics for Class 1 are more important because this class is associated with underrepresented severe convective phenomena. The last two rows represent metrics for the aggregation stage. Lower values of the latter imply the ability of the model to capture the annual cycle of SCS. The support stands for the number of elements in subsets corresponding to Class 1 and Class 2.}
\begin{center}
\begin{tabular}{p{60mm}|lll|lll|l|l}
 \multirow{2}{*}{Model}          & \multicolumn{3}{c|}{\textbf{Class 0: no SCS}} & \multicolumn{3}{c|}{\textbf{Class 1: SCS}} &   
\multirow{2}{*}{RMSE} &
\multirow{2}{*}{RMSE$_{AC}$}\\

&Precision&Recall&F1&Precision&Recall&F1&&\\
\hline


CatBoost baseline   &    0.82      &    0.34     &    0.48     &    0.31     &    0.79     &     0.40  & 0.0100 & 0.0121  \\

Autoencoders \& classifier (separate)&  0.82 & 0.53 & 0.64 & 0.32 & 0.66 & 0.43 & 0.0086 & 0.0138\\

Autoencoders \& classifier (end-to-end) & 0.83 & \textbf{0.54} & \textbf{0.65} & 0.33 & 0.67 & 0.44 & \textbf{0.0080} & 0.0136\\

Stacked autoencoders with  summation \&   classifier & 0.83 & 0.53 & 0.65 & \textbf{0.34} & 0.69 & \textbf{0.45} & 0.0084 & 0.0125   \\

Stacked autoencoders with conv1d layer \& classifier  &   0.84  &   {0.50}   &  {0.63}   &   {0.33}  &  0.71   & \textbf{0.45} & {0.0093} & 0.0154  \\ 

Stacked autoencoders with   linear layer \& classifier  &   \textbf{0.85}  & 0.38 & 0.53 & 0.31 &  \textbf{0.81}   & \textbf{0.45} & 0.0125 & \textbf{0.0114}  \\ \hline

    Support       & \multicolumn{3}{c|}{68015} & \multicolumn{3}{c|}{23360} &   \multicolumn{1}{c|}{} \\ 
\end{tabular}
\label{tab: metrics}
\end{center}
\end{table*}

Although incremental improvement and expansion of the main model yield marginal gains, a threshold cannot be surpassed through architecture modifications alone.

We believe this threshold may be attributed to the distinct nature of the datasets being used. Meteorological observations come from the real world, while CMIP5 data are climate models. They do not accurately represent the real state of the atmosphere in the far future but rather simulate the trend of its change given a particular RCP scenario.

However, it is worth noting that during the CMIP5 model quality evaluation process, all authors simulate climate for a given historical period. Their model is only approved if it matches the actual climate change. Therefore, verification of CMIP5 datasets by their authors and the daily aggregation of features we use during preprocessing help the model to perform reasonably well.

\section{Discussion}


We also see a narrow extended area of activity in the middle of the picture, which is associated approximately with the Southern part of the Ural Mountains. The Ural Mountains, with their rugged terrain, also contribute to orographic lifting and the development of severe convection.    
\begin{figure}[htbp]
\centerline{\includegraphics[width=10cm,height=9cm,keepaspectratio]{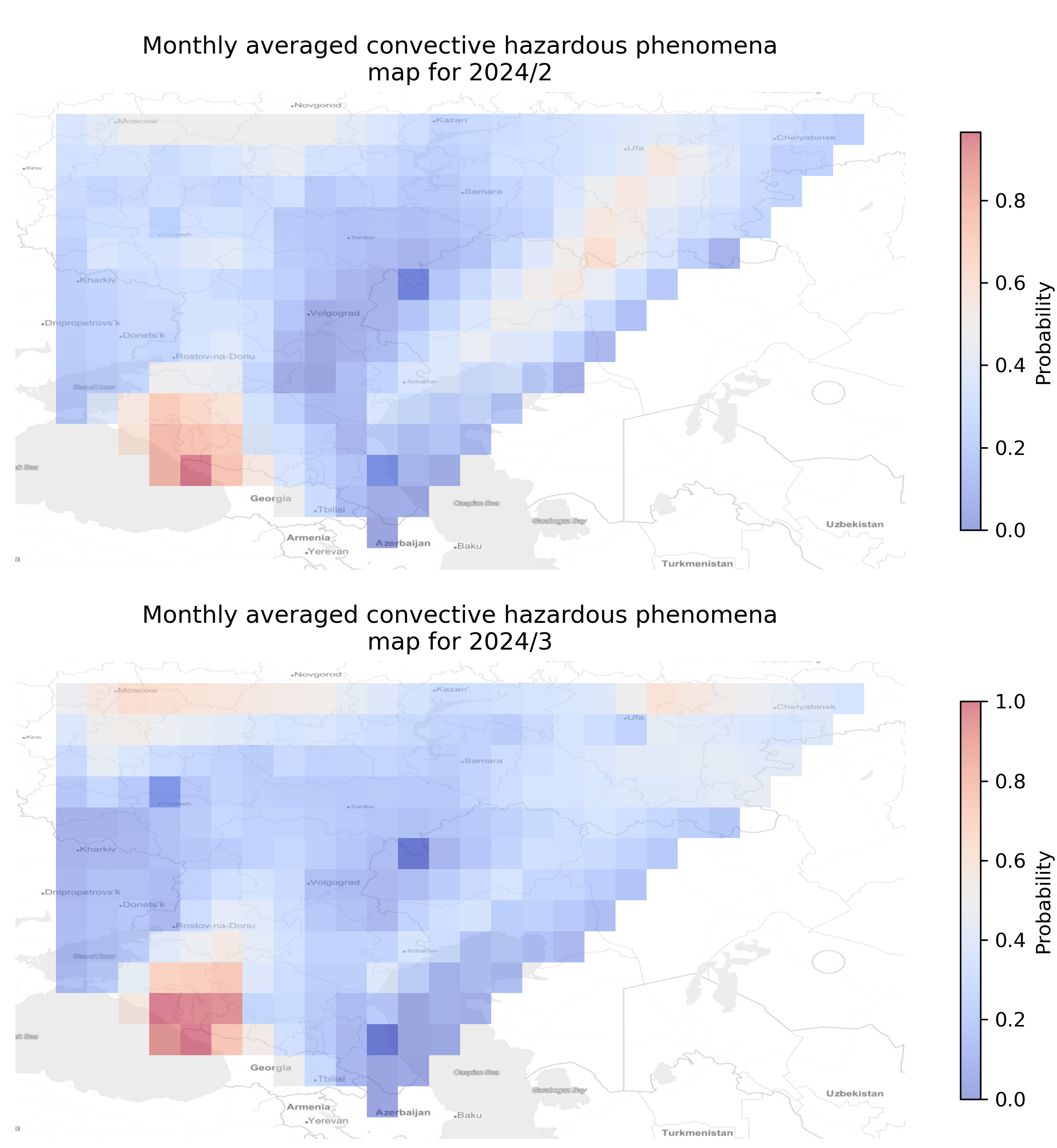}}
\caption{The monthly averaged forecasted SCS probabilities for February and March (top to bottom) for Southern territories of Russia. Areas of high SCS activity are depicted in red.}
\label{fig:winter}
\end{figure}

Figure~\ref{fig:winter} represents the forecast for February (top) and March (bottom) 2024. The figure suggests there may be areas of high SCS frequency in the Black Sea shore - the Caucasus region. This finding is valid and significant based on an analysis of historical climate data. Here, the mountainous terrain of the Caucasus region provides favorable conditions for severe convection. The mountains cause orographic lifting, leading to cloud formation and enhanced convection.

Additionally, the sharp elevation changes in the region create localized areas of instability, contributing to the development of severe thunderstorms. The proximity to the Black Sea results in necessary atmospheric moisture for convection in this area. The latter is the most active area in the forecasting scope. 

We also see a narrow extended area of activity in the middle of the picture, which stands for the Southern part of the Ural Mountains. The Ural Mountains, with their rugged terrain, also contribute to orographic lifting and the development of severe convection.    
\begin{figure}[htbp]
\centerline{\includegraphics[width=10cm,height=9cm,keepaspectratio]{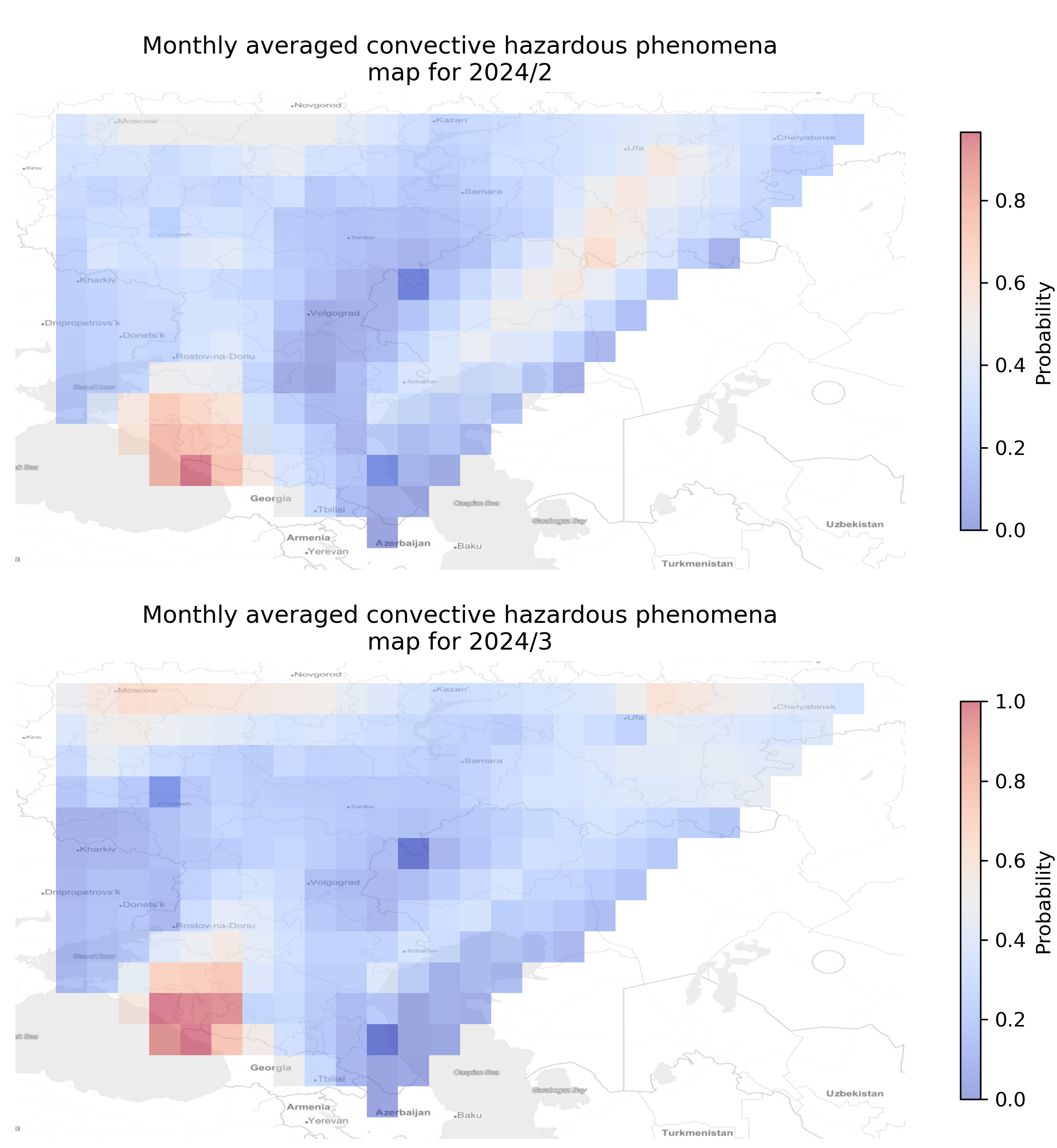}}
\caption{The monthly averaged forecasted SCS probabilities for February and March (top to bottom), the Caucasus region. The areas of high SCS activity are in red.}
\label{fig: winter}
\end{figure}

\begin{figure}[htbp]
\centerline{\includegraphics[width=10cm,height=9cm,keepaspectratio]{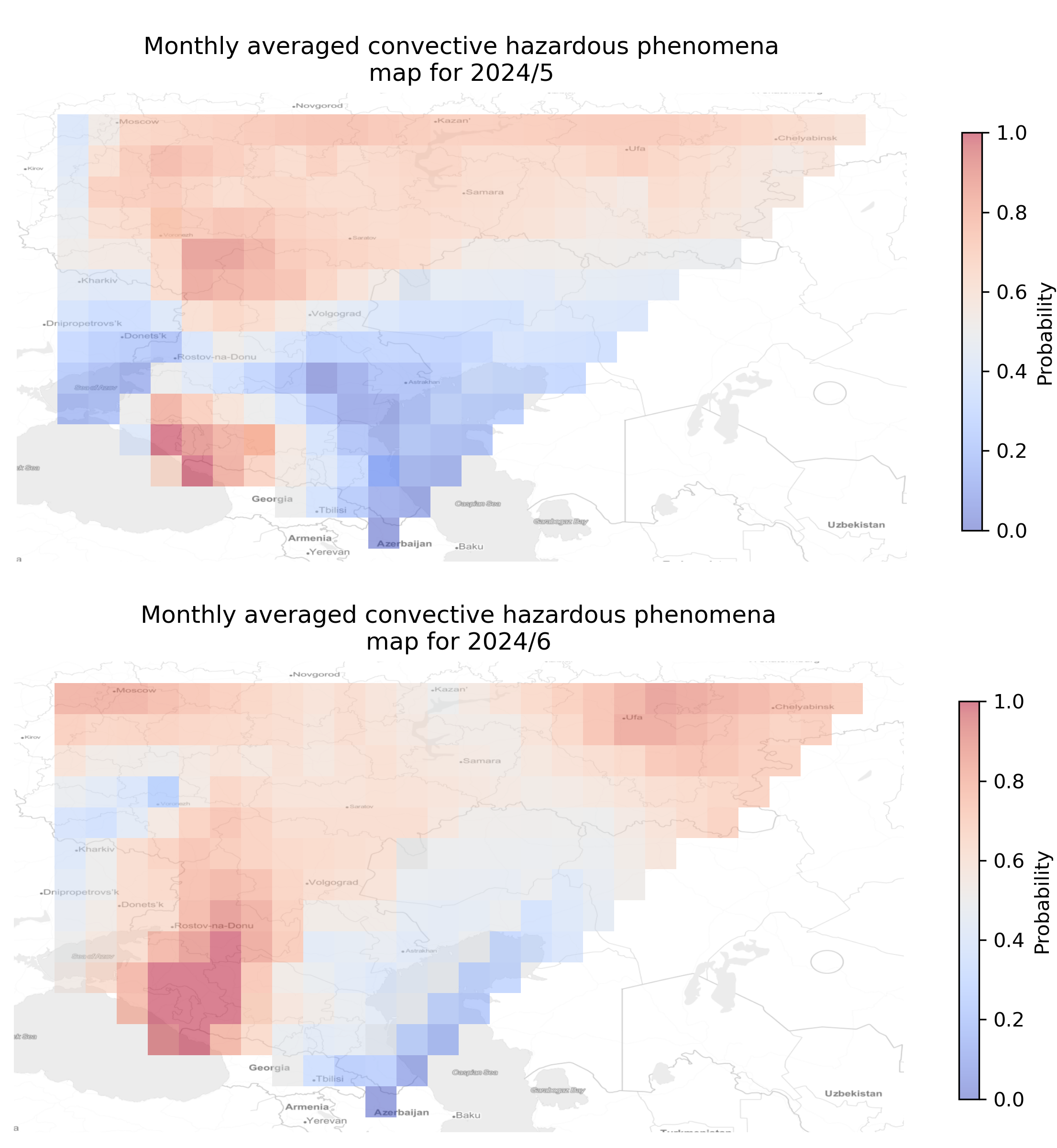}}
\caption{The monthly averaged forecasted SCS probabilities for May and June (top to bottom) for the Caucasus region. Areas of high SCS activity are in red.}
\label{fig: summer}
\vspace{-2mm}
\end{figure}

Figure \ref{fig: summer} shows the forecast for May (top) and June (bottom) of 2024.
During the summer, there is an increase in the size of areas occupied by SCS-active areas and the frequencies of SCS in the North-East from the Black Sea.

\begin{figure}[htbp]
\centerline{\includegraphics[width=10cm,height=9cm,keepaspectratio]{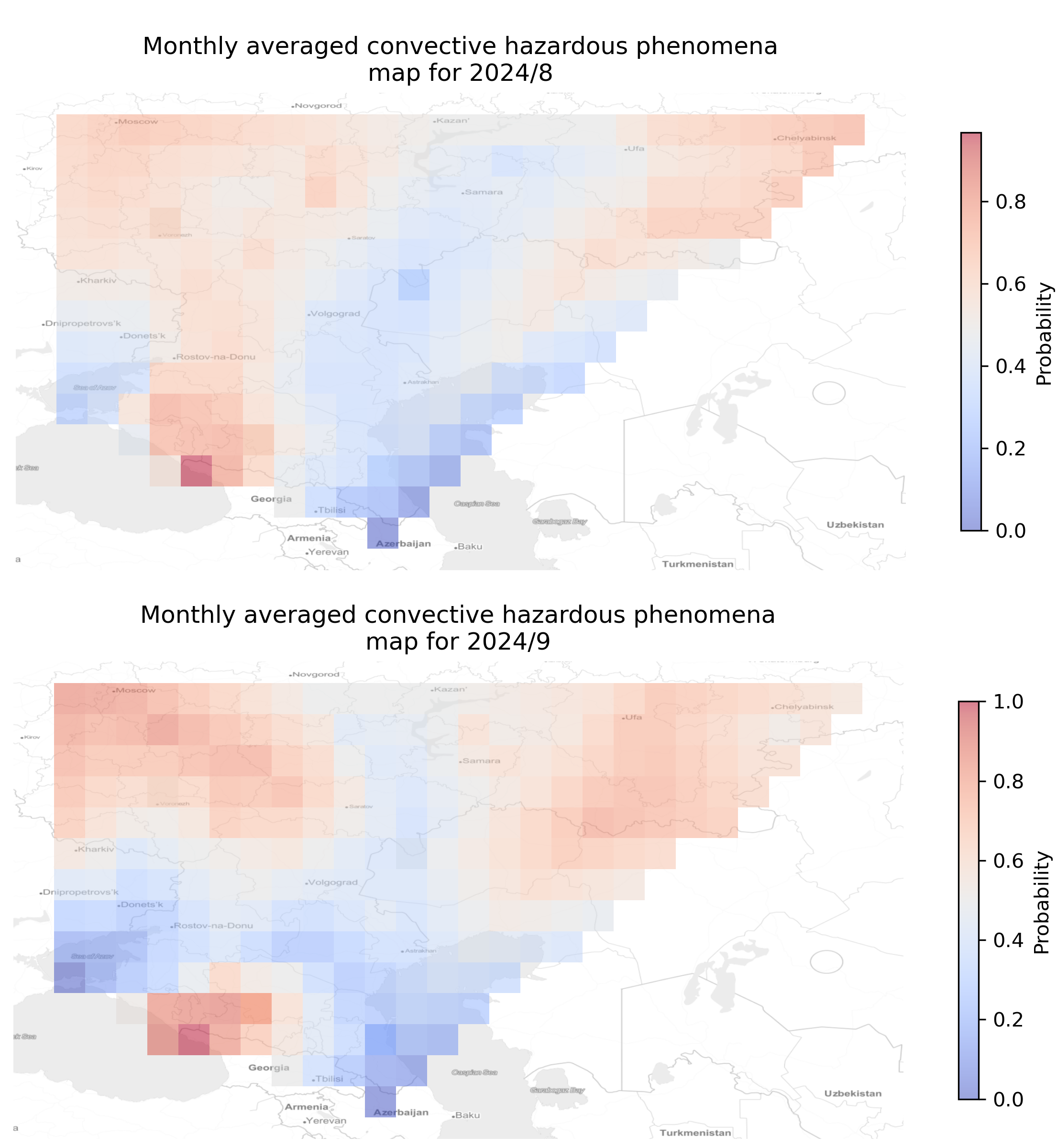}}
\caption{The monthly averaged forecasted SCS probabilities for August and September (top to bottom) for the Caucasus region. Areas of high SCS activity are in red.}
\label{fig: autumn}
\vspace{-2mm}
\end{figure}

Figure \ref{fig: autumn} represents the forecast for August (top) and September (bottom), 2024. In the late summer to early autumn, areas of high activity remain relatively extensive but exhibit a smoother distribution. Also, peak activity regions along the Black Sea shore are not too far from the other areas.

\section{Conclusion}

This paper introduces a new model designed to produce climatological SCS frequency forecasts based on CMIP data. We tested several architectural modifications, including the autoencoder hourglass stack \cite{newell2016stacked} and different approaches to embeddings and reconstructions aggregation. We compared our model to the CatBoost baseline. Our model significantly outperforms it.
Also, we presented a sample SCS forecast for the European part of Russia and conducted a brief analysis of the results.

Still, there are limitations to our model as it is tailored to the specific locations it has been trained on, which means that retraining and adaptation to the target dataset are necessary when using it for different regions. Additionally, the model tends to be overly cautious, sometimes labeling non-SCS events as SCS. This is reflected in a high recall but relatively low precision, indicating that while the model can accurately identify most SCS events, it generates many false positives. When minimizing the risks associated with SCS for a business, false positive errors are preferable to false negative ones.

Considering the inherent limitations, our model contributes to the field of SCS risk estimation. Assessing changes in SCS frequency might help to reduce costs for dependent businesses and governmental organizations, such as agricultural and insurance companies, as well as the green energy sector (especially regarding solar panel installations). One may use the prediction outputs to enhance modern economic risk models.

We would like to emphasize that with the existing datasets, researchers need to exercise caution, as a single dataset, such as the Storm Events Database, may be compiled from diverse sources, introducing significant data inhomogeneity in some instances. Moreover, observation techniques and the overall quality of observations improve over time, leading to an artificial increase in the SCS trend. Hence, it is crucial to consider these factors when working with historical data.

Nevertheless, the modernization of meteorological stations and the expansion of station coverage will undoubtedly yield substantial benefits for SCS forecasting and advice to communities and stakeholders regarding the optimal adaptation measures.

We plan to continue our work and are working with the witness-based dataset of dangerous atmospheric phenomena called the European Severe Weather Database, provided by the European Severe Storms Laboratory. This dataset primarily covers European countries and includes data from several cities.

\bibliographystyle{unsrtnat}
\bibliography{references}

\end{document}